\documentclass [12pt,fullpage,amssymb,twoside]{article}
\usepackage{psfig}
\usepackage{bbbold}
\def\slantfrac#1#2{\hbox{$\,^{#1}\!/_{#2}$}}
\def\mdot{{\raisebox{1pt}{\hbox{$\stackrel{\bullet}{M}$}}}\ \!\!}
\def\apgt{\ {\raise-.5ex\hbox{$\buildrel>\over{\scriptstyle\sim}$}}\ }
\def\aplt{\ {\raise-.5ex\hbox{$\buildrel<\over{\scriptstyle\sim}$}}\ }
\def\N{\symbol{242}}
\begin{document}

\title{Gas Flow Structure in Binary Systems with Mass Exchange Driven by
Stellar Wind}

\author{A.A.Boyarchuk$^1$, D.V.Bisikalo$^1$, E.Yu.Kilpio$^1$,
O.A.Kuznetsov$^{1,2}$\\[5mm]
\small
$^1$ Institute of Astronomy RAS, Moscow\\
\small $^2$ Keldysh Institute for Applied Mathematics, Moscow, Russia\\}

\date{}

\maketitle

\begin{abstract}

Results of calculations of the flow structure in binary systems with
mass exchange driven by stellar wind are presented. 2D simulations
have been carried out using the Roe-Oscher scheme. The fine grid we
used allowed us to detect the details of the flow structure. In
particular, it was found out that solutions with the wind velocity $V$
of the order of the orbital velocity of the system $V_{s}$ are
unstable: in solutions with $V<V_{s}$ -- the steady accretion
disk takes place and if $V>V_{s}$ -- the cone shock forms.
Consideration of the minor variations of the wind velocity $V$ around
$V_s$ shows that they can influence greatly the flow structure
causing transition from disk accretion to the accretion from the
flow. During the flow rearrangement period (the accretion disk
destruction) the accretion rate increases in dozens times for a
short time ($\sim$ 0.1 of the orbital period of the system). This
change can result in the drop-off of the gas from the accretor that
is usually associated with activity in these systems.

\end{abstract}


\section{Introduction}

The overwhelming majority of stars are binaries. In wide pairs the
effect of one component to another is negligible but if the
components are close enough we can't ignore the interaction between
them. One of the most bright manifestations of this interaction is
the mass exchange between the components. Such systems are called
interacting binaries. The study of the flow structure in these
systems is important for the consideration of their evolutionary
status as well as for interpretation of observations.

Surely, the strong interaction is observed for close systems
where at least one of components fills its critical surface (so
called Roche lobe), i.e. equipotential surface passing through the
inner Lagrangian point $L_1$ (see, e.g., [\ref{Kniga}]). In the $L_1$
point the pressure gradient is not balanced by the gravitational
force and the component that fills its Roche lobe loses matter
through the
$L_1$ point.  But in many cases even if no one of components fills
its Roche lobes the mass transfer between them may be rather
intense due to the stellar wind from one of the components (or in
some cases from both of them). In this work we consider the flow
structure in a binary system with mass exchange driven by stellar
wind using symbiotic stars as the example.

Now it is generally recognized that classic symbiotic stars are binary
systems consisting of a cool giant and a hot compact star (white dwarf
or subdwarf) surrounded by nebulosity. Symbiotic stars are also characterized
by a significant photometric and spectroscopic variability and,
in particular, by the irregular flare activity. It is also avowed that
components in these systems don't fill their Roche lobes and there are
evidences that the giant loses mass at the high rate due to stellar wind
and the intense interaction between the components takes place. These
hypotheses are also supported by observations (see, e.g., [\ref{Boyar92}]).

In order to explain the flare activity in symbiotic stars different
mechanisms were proposed (see, e.g., [\ref{MK92}]). In particular, the most
probable scenario for a classic symbiotic star is the one where the additional
energy release is due to the fluctuations of the accretion rate near the level of
the stationary thermonuclear burning [\ref{FC95}]. If the accretion rate exceeds the
maximum value when the hydrogen can burn in a layer source of a degenerate
core, the accreting gas stores above the burning layer and grows to
the giant's size [\ref{TU}--\ref{IbenTu}]. This process is usually
associated with outbursts in classical symbiotic stars. What are these
changes in the accretion rate caused by? In a number of works
(e.g., [\ref{Boyar92},\ref{Friedjung93}]) some attempts to explain such a
flare activity by the giant's wind variability were made but the observations
carried out before the outburst registered only insignificant changes of the wind
[\ref{Friedjung93},\ref{Kenyon83}]. This fact imposes restrictions to all
outburst scenarios caused by changes of giant's characteristics.

In this work we have carried out the 2D numerical investigation of
mass transfer in classical symbiotic systems using TVD method that
is considered to be the best for numerical simulation of mass
transfer in binary systems~[\ref{Kniga}]. The main attention was
paid to consideration of the solution where stellar wind velocity has
changed from below to above of the orbital velocity of the system
as well as to study of the transition from the disk accretion to the
accretion from the flow and appropriate rearrangement of the flow
structure.

\section{Model}

All the calculations were carried out for a binary system with parameters of the classic
symbiotic star -- Z~And.

In order to study the gas flow structure in the equatorial plane of a classical symbiotic
system the 2D numerical simulation has been carried out. The zero point of the coordinate
system was placed to the center of the donor, $x$ axis was directed along
the line connecting centers of the components, $y$ axis -- along the accretor's orbital motion.
The flow was described by the system of Euler's equations in the coordinate frame rotating with the angular
velocity of the binary system $\Omega $:

\[
{\frac{{\partial \rho} }{{\partial t}}} + {\frac{{\partial \rho
u}}{{\partial x}}} + {\frac{{\partial \rho v}}{{\partial
y}}}=0\,,
\]

\[
{\frac{{\partial \rho u}}{{\partial t}}} + {\frac{{\partial (\rho \,u^{2} +
p)}}{{\partial x}}} + {\frac{{\partial \rho uv}}{{\partial y}}} = - \rho
{\frac{{\partial\Phi} }{{\partial x}}} + 2\Omega v\rho\,,
\]

\[
{\frac{{\partial \rho v}}{{\partial t}}} + {\frac{{\partial \rho
uv}}{{\partial x}}} + {\frac{{\partial (\rho \,v^{2} + p)}}{{\partial y}}} =
- \rho {\frac{{\partial\Phi} }{{\partial y}}} - 2\Omega
u\rho\,,
\]

\[
{\frac{{\partial \rho E}}{{\partial t}}} + {\frac{{\partial \rho
uh}}{{\partial x}}} + {\frac{{\partial \rho vh}}{{\partial y}}} = - \rho
u{\frac{{\partial \Phi} }{{\partial x}}} - \rho v{\frac{{\partial \Phi
}}{{\partial y}}}\,.
\]
Here ${\bmath u}=(u,v)$ is the velocity vector, $p$ -- pressure,
$\rho$ denotes density,
$h=\varepsilon+{{p}\mathord{\left/{\vphantom {{p}{\rho}}}
\right.\kern-\nulldelimiterspace}{\rho}}+{{{\left|{\bmath u}
\right|}^{2}}\mathord{\left/{\vphantom{{{\left|{u}
\right|}^{2}}{2}}}\right.\kern-\nulldelimiterspace} {2}}$ --
specific total enthalpy, $E=\varepsilon+{{{\left|{\bmath
u}\right|}^{2}}\mathord{\left/ {\vphantom {{{\left|{u}
\right|}^{2}}{2}}}\right.\kern-\nulldelimiterspace}{2}}$ --
specific total energy, $\varepsilon$ -- specific intrinsic energy
,
$\Phi({\bmath r})$ --  force potential.

In the standard problem definition when only gravitational forces
from the point mass components and centrifugal force are taken into
account the force potential looks as the following:

\[
\Phi({\bmath r})=-\frac{GM_1}{|{\bmath r}-{\bmath r}_1|}
-\frac{GM_2}{|{\bmath r}-{\bmath r}_2|}
-\slantfrac{1}{2}\Omega^{2}({\bmath r}-{\bmath r}_c)^2\,.
\]
Here $M_1$ is the donor's mass, $M_2$ -- mass of the accretor,
${\bmath r}_1$, ${\bmath r}_2$ are the radius-vectors of
the centers of components, ${\bmath r}_c$ -- radius-vector of the
mass center of the system. This is so-called Roche potential. But in
our case the additional force responsible for donor's wind
acceleration should be also taken into account. Therefore, the form
of the potential will change.

Previous studies (e.g., [\ref{GS}--\ref{Dima96}]) have shown that
the general flow structure in the system where components do not
fill their Roche lobes is defined first and foremost by the stellar
wind parameters. Unfortunately, wind velocity regime is not
well-known  due to the absence of an avowed mechanism of gas
acceleration in stellar atmospheres, so we used the parametric
representation for the force responsible for the acceleration of the
wind in the following form:

\[
{\bmath F}({\bmath r}) =
\alpha\frac{GM_1}{|{\bmath r}|^{2}}\cdot\frac{\bmath r}{|{\bmath
r}|}\,.
\]
where $\alpha$ -- parameter.

After taking into account this force directed from the donor,
the modified force potential looks as the following:

\[
\Phi({\bmath r})=-\frac{GM_1}{|{\bmath r}-{\bmath r}_1|}
+\alpha \frac{GM_1}{|{\bmath r}-{\bmath r}_1|}
-\frac{GM_2}{|{\bmath r}-{\bmath r}_2|}
-\slantfrac{1}{2}\Omega^{2}({\bmath r}-{\bmath r}_c)^2\,.
\]

Results of numerical modelling carried out for a wide range of $\alpha$
(see [\ref{Lena}]) have shown that if $\alpha<0.8$ there is no stationary
solution and gas injected to the space between the components falls back
to the donor's surface. Finally there are no matter left to form the
circumbinary envelope (the obtained densities do not exceed the background
value). The situation is rather different when $\alpha$ has larger values.
In these cases stationary regime takes place after few orbital periods from the
beginning of calculations. In this work we accept $\alpha=1$ as
in [\ref{Theuns1},\ref{Theuns2}], i.e. we assume that the accelerating force
balances the donor's gravitational force.
The accepted value of $\alpha$ provides the wind velocity change according to
the $\beta$-law (Lamers law) with $\beta\approx1$ and the value at infinity
$V_\infty$ that are in agreement with observations.

To complete the system the ideal gas equation of state was used

\[
p = (\gamma-1)\rho \varepsilon\,,
\]
where the ratio of the specific heats $\gamma$ was accepted to be
$\slantfrac{5}{3}$. The solution with $\gamma=\slantfrac{5}{3}$
corresponds to the case without energy losses and gives correct results
only for a system where radiative losses are negligible, i.e. for an
optically thick media. The observed value H/K of C{\,\sc iv}
1550{\AA} doublet shows that in a quiescent state the circumbinary
envelope in Z And is optically thin~[\ref{FC95}]. We don't consider
all the envelope but only the area in the vicinity of the accretor
in the equatorial plane of the system, namely the disk. As our
estimates show, in the case of Z~And the observed values of
accretion rate is $\mdot_{accr}=4.5\cdot10^{-9}M_\odot$/year and the
disk with $T=15000$~K will be optically thick i.e. in the area we
study the use of the adiabatic model is correct. The optical thickness $\tau$
is defined as the product of the absorption coefficient $\kappa$,
density and the layer's width $\tau=\kappa\cdot\rho\cdot l$. In the
case of disk accretion the ratio of the layer width where $\tau=1$
to the disk width is the pacing factor. For the case of disk
accretion

$$
\mdot_{accr}=2\pi R\cdot H\cdot\rho v_r\,,
$$
where $R$ -- disk radius, $H$ -- disk half width,
$v_r$ --  radial component of the velocity

$$
v_r=\frac{\alpha_{ss} c_s H}{R}\,,
$$
where $c_s$ -- sonic velocity, $\alpha_{ss}$ -- Shakura-Suynyaev parameter,
and

$$
H=\frac{c_s}{V_K}R\,,
$$
where $V_K$ is the Keplerian velocity, and finally this
factor can be expressed as:

$$
\frac{l^{\tau=1}}{H}
=\frac{2\pi\alpha_{ss}}{\kappa\mdot_{accr}}\cdot\frac{c_s^2\cdot R}{V_K}
=\frac{2\pi\alpha_{ss}}{\kappa\mdot_{accr}}\cdot
\frac{{\cal R}T\cdot R^{3/2}}{\sqrt{GM}}\,,
$$
where $\cal R$ is the gas constant.
For the typical disk temperatures $T\simeq15000$~K the absorption
coefficient depends on density weakly and equals $\simeq$ 100
cm$^2$/g [\ref{BellLin}]. Let us assume $T=15000$~K,
$R=50R_\odot=A/10$,
$M=0.6M_\odot$ (in this case $c_s=11$~km/s, $V_K=49$~km/s, $H=11R_\odot$)
then we will obtain:

$$
\frac{l^{\tau=1}}{H}=
8\cdot10^{-3}
\cdot\left(\frac{\alpha_{ss}}{0.1}\right)
\cdot\left(\frac{\kappa}{100~\mbox{cm}^2/\mbox{g}}\right)^{-1}
\cdot\left(\frac{\mdot_{accr}}{10^{-8}M_\odot/\mbox{yr}}\right)^{-1}\,,
$$
So, when $\mdot_{accr}=4.5\cdot10^{-9}M_\odot$/yr the value of the optical
width reaches the value $\tau=1$ at less than 2\% of the disk's width and
we can conclude that the solution with $\gamma=\slantfrac{5}{3}$ is
applicable for considered systems.

Parameters of Z~And we used in our calculations were taken from
[\ref{FC88}] and are summarized in the Table~1. The choice of namely
Z~And for our calculations is rather reasonable because Z~And is one
of the most studied classic symbiotic stars. The studies of the
energy distribution in the wide spectral range (from UV to IR)
during the quiescent period (1978-1982) [\ref{FC88}] allowed to
estimate the characteristics of Z~And: the donor (cool M3.5III
giant) loses mass at the rate of $2\times10^{-7}~M_\odot$/yr; the
hot compact component (accretor) has the temperature $\sim 10^5$~K
and accretes the small part of the gas of the donor's wind (about
2\%). Gas of the circumbinary envelope has the electron density of
$\sim2\cdot10^{10}~\mbox{cm}^{-3}$ and the temperature
$\sim1.5\div8\cdot10^4$~K. Systematic observations of Z~And allowed
to estimate its characteristics during the outburst and to find the
features that can be easily explained in terms of the formation and
subsequent drop-off of the optically thick shell by the accretor at
the rate of about 250--300~km/s.

\begin{table}[h]
\begin{center}
Table 1. Parameters of Z~And \\[5mm]
\begin{tabular}{|l|c|}
\hline
Parameter&Value\\
\hline\hline
Mass of the donor $M_1$&$2M_\odot$\\
\hline
Radius of the donor $R_1$&$77R_\odot$\\
\hline
Mass of the accretor $M_2$&$0.6M_\odot$\\
\hline
Radius of the accretor $R_2$&$0.07R_\odot$\\
\hline
Orbital period of the system $P_{orb}$&760~days\\
\hline
Separation $A$&$483R_\odot$\\
\hline
Radius of the donor's Roche lobe&$232R_\odot$\\
\hline
Orbital velocity of the donor&7~km/s\\
\hline
Orbital velocity of the accretor&25~km/s\\
\hline
\end{tabular}
\end{center}
\end{table}

To solve the system of the 2D gasdynamic equations
the TVD-type Roe scheme [\ref{Roe}] with the restrictions of
fluxes in the Oscher form [\ref{Osher}] was used. The scheme is
quasi-monotonic and has the third order of spatial accuracy and
the first order of time accuracy. The uniform grids consisting
of $301\times301$ nodes and $601\times601$ nodes were used.
The considered domain has the form of a square $[-A\ldots
2A]\times[-\slantfrac{3}{2}A\ldots\slantfrac{3}{2}A]$ with excluded
circles of radii equal to the ones of components and centers in
the component's centers. The free outflow boundary conditions
($\bmath u=0$, $p=0$) were accepted on the outer border as well as
on the accretor surface. On the surface of the donor star the
boundary conditions of the mass inflow were adopted. It should be
mentioned that the boundary value of the density on the donor's
surface doesn't influence the solution because the set of equations
used allows scaling of the
$\rho$ (with the simultaneous scaling of $p$). Boundary conditions
on the donor are following: density $\rho(R_1)=1$, temperature
$T(R_1)=3200$~K, the gas is injected to the system along the vector normal
to the surface and its velocity is equal to $V$. The simulations were
carried out in the velocity $V$ range from 25 to 75 km/s.
The solving of the set of equations was carried out by means of
method of establishment from initial state with parameters
$\rho_0=10^{-5}\rho(R_1)$;
$p_{0}=10^{-4}\rho(R_1)\,c^2(R_{1})/\gamma$, ${\bmath u}_0=0$
up to steady state configuration.

\section{Results}

\begin{figure}
\centerline{\hbox{\psfig{figure=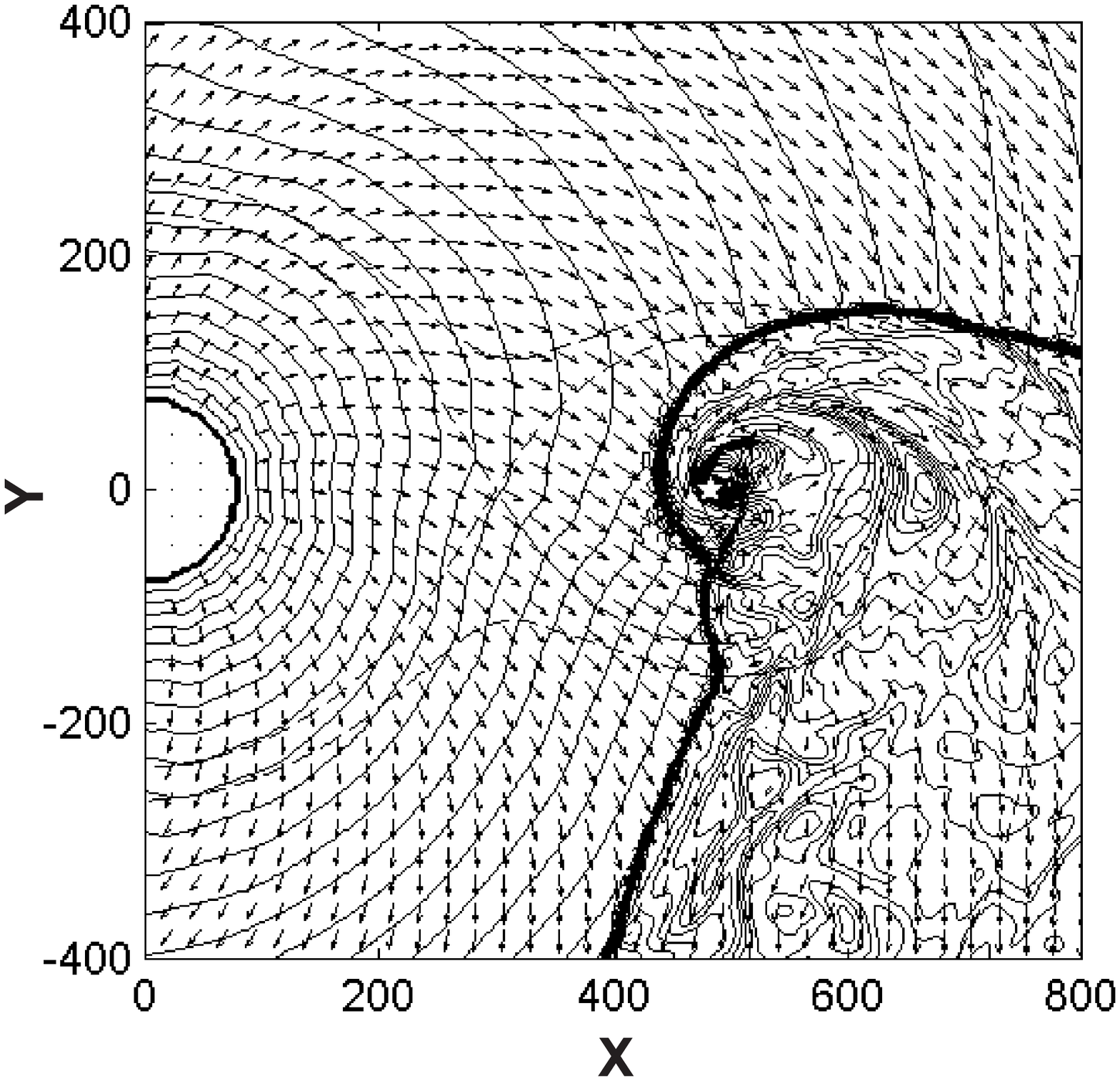,width=9cm}}}
\centerline{\hbox{\psfig{figure=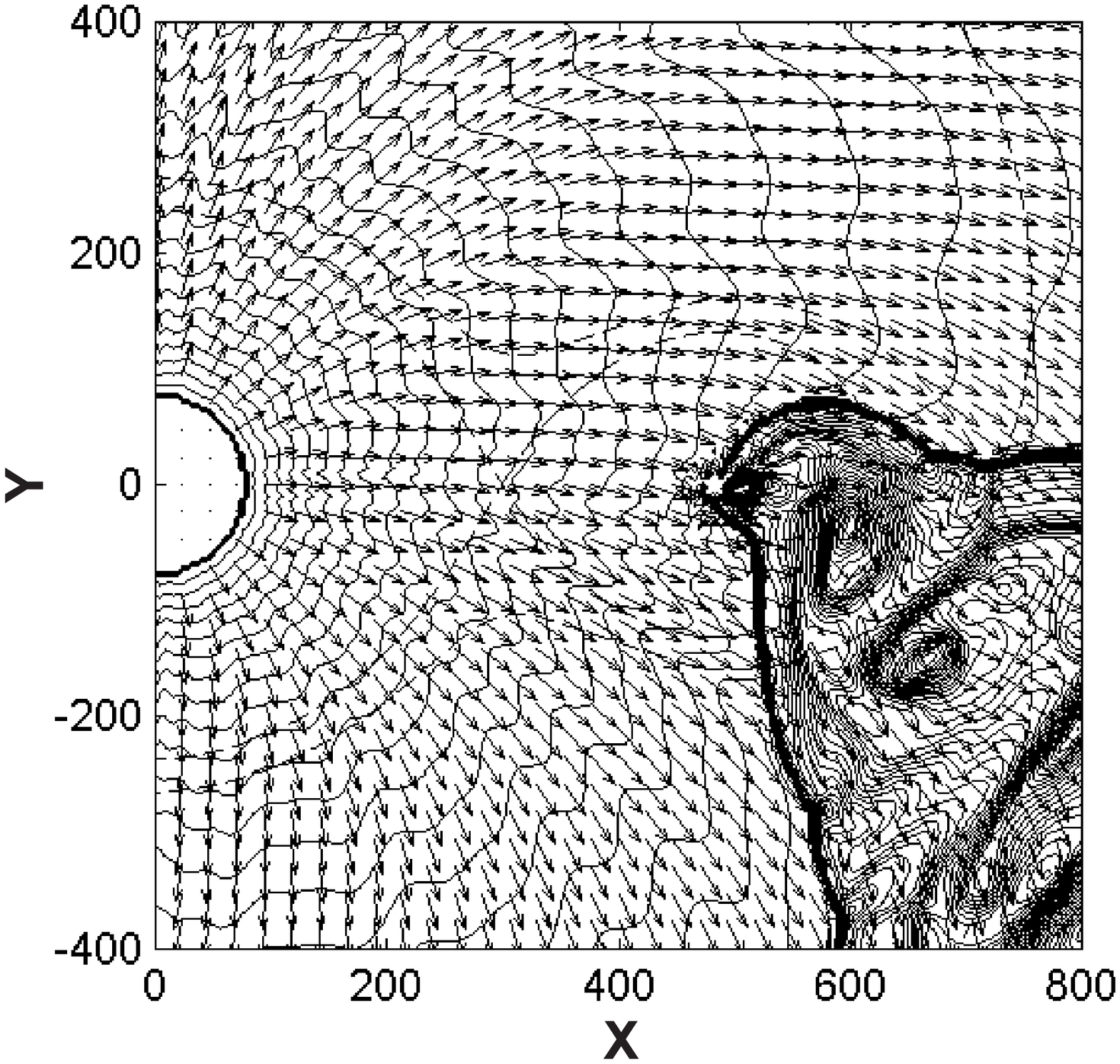,width=9cm}}}
\caption{\small{\it Top panel:} Density field and velocity vectors
for the stationary case $V=25$~km/s. The isolines of the density
logarithm in the range of $10^{-5}$ to 1 (in the units of density at
donor's surface) are shown. The distances are given in solar radii.
Dashed lines show equipotential contours of the standard Roche
potential. Donor's surface is shown by the half circle with the center
in (0,0) point and the radius $R_1$. The star symbol marks accretor.
 {\it Bottom panel:} The same as in the upper panel for the case $V=75$~km/s.
Density contours in the bottom panel correspond to density values in the range of
0.04 to 0.55 (in the units of density at the donor's surface).}
\label{regimes}
\end{figure}

The results have shown that when the wind velocity is less than
$V=35$~km/s the bow shock and the accretion disk behind it form in
the system and the stationary gas flow regime takes place
(see the top panel of Fig.~\ref{regimes}).
We consider the stationary regime to be reached when the amount of gas
injected to the system becomes equal to the one that leaves it
(part of the gas is being accreted and part leaves the considered
domain through the outer border). After that the total amount
of the gas in the system doesn't change and the accretion rate
remains constant. It happens after approximately four orbital
periods after beginning of calculations. Results of calculations
show that the greater the wind velocity the less the distance
between the bow shock and the accretor and when the value
of velocity is big enough the disk
disappears and the shock becomes the attached one so the disk
does not form (see the bottom panel of Fig.~\ref{regimes}). The accretion rate in
these cases is slightly higher
[\ref{AZh2002}].

The observed values of the wind velocity in symbiotic stars are not
far from the value $V=35$~km/s that divides two principally
different cases with different accretion regimes. So if at some
moment the wind's velocity increases and exceeds 35 km/s the the
disk will be destroyed and the accretion regime will change. It
should be mentioned that the value $V=35$~km/s dividing different
accretion regimes is also close to the orbital velocity (32 km/s).
This fact is in accordance with [\ref{Dima96}].

In order to consider the process of transition between these two
regimes after the wind velocity increase we took the stationary
solution for the case
$V=25$~km/s, increased the velocity up to $V_1$ and continued calculations.
Different cases with $V_1$ values in the interval 35~km/s~$\le V_1\le75$~km/s
were considered.

\begin{figure}
\vspace*{5mm}
\centerline{\hbox{\psfig{figure=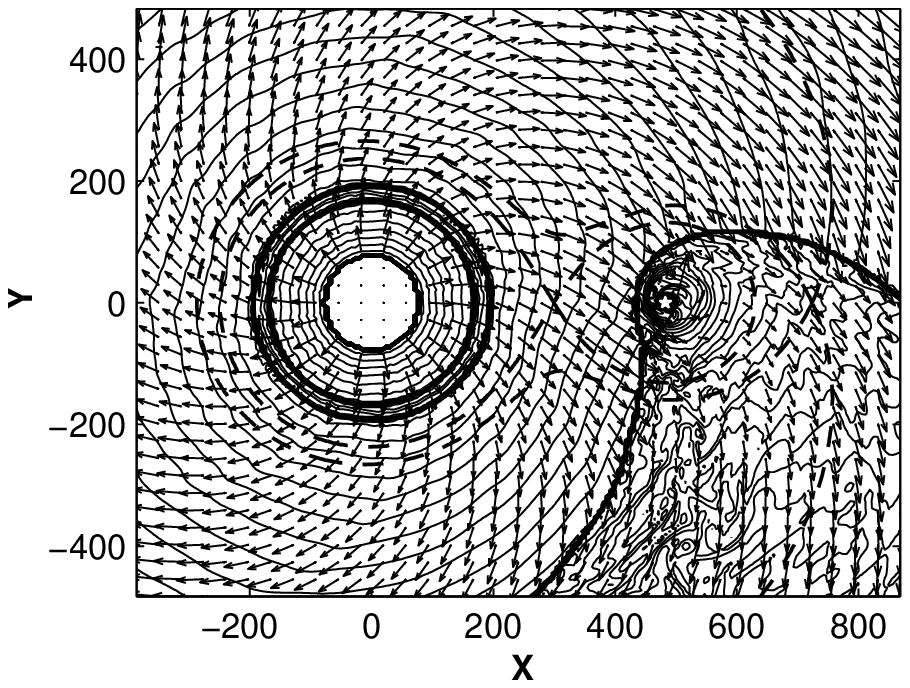,width=10cm}}}
\vspace*{3mm}
\centerline{\hbox{\psfig{figure=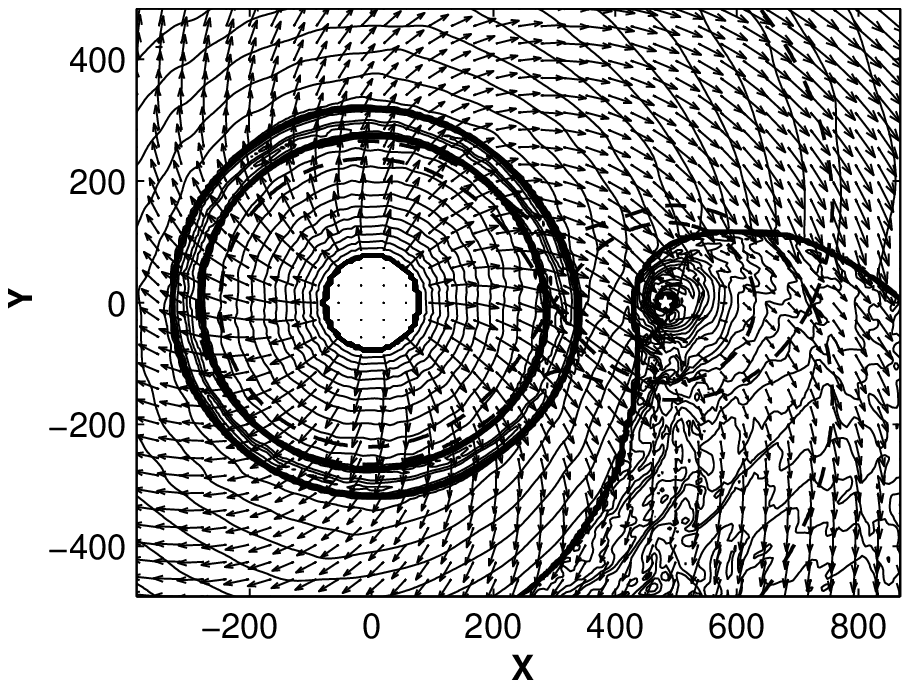,width=10cm}}}
\vspace*{5mm}
\caption{\small Density field and velocity vectors for the case  after wind
velocity change from $V=25$~km/s to $V_1=40$~km/s.
Isolines of the density logarithm in the range of $10^{-5}$ to 1
(in the units of density at donor's surface) are shown. Distances
are given in solar radii. Dashed lines show the equipotential contours of the
standard Roche potential. The star symbol marks accretor.
{\it Top panel:} the moment $0.034P_{orb}$ after the wind velocity change.
{\it Bottom panel:} the moment $0.073P_{orb}$ after the wind velocity change.}
\label{double_shock}
\end{figure}

In Figure~\ref{double_shock} the morphology of the gas flow -- density field and
velocity vectors (shown by arrows) -- for the case of the wind
velocity change from $V=25$~km/s to $V_1=40$~km/s is presented.
Donor's surface is shown by the circle with the center
in (0,0) point and the radius $R_1$. Dashed lines mark the standard
Roche equipotential contours. The star symbol marks accretor. Distances are
given in the units of the solar radius. Two moments are
presented: the first one is for $0.015P_{orb}$ (top panel) and the
second one is for
$0.05P_{orb}$ (bottom panel) after the wind velocity change. We can
see that after the wind velocity change two shocks (seen as
condensations of density isolines) propagating from the donor's
surface are formed. Being very close to each other at the first moments
they begin to diverge gradually with time. Let us try to explain
this phenomenon. Changing of the velocity regime on the surface of
mass-losing star results in the collision of two codirectional
flows: the `new' one (moving with the radial velocity 40~km/s) and
the `old' one (25~km/s). This collision can be described in terms of
the Riemann problem. Introducing a new coordinate system moving with
the velocity of 32.5 km/s in the radial direction it can be reduced
to the frontal collision of two flows with velocities $\pm$7.5~km/s.
The solution of this Riemann problem is two diverging shocks moving
with velocities (see, e.g., [\ref{Samarskii}])

$$
D_{1,2}=\pm\left(\frac{\gamma+1}{4}U
+\sqrt{\left(\frac{\gamma+1}{4}U\right)^2+c_s^2}-U\right)\,.
$$
Substituting $\gamma=\slantfrac53$, $U=7.5$~km/s, $c_s=5$~km/s
we obtain $D_{1,2}=\pm4.5$~km/s. Coming back to the original
coordinate system we obtain two successive shocks
with velocities $D_1=37$~km/s, $D_2=28$~km/s. Note that these
values correspond to the first moments of time after the
switching of the velocity regime and later will be changed under
the action of forces and variation of $c_s$.

Let us also estimate the strength of these shocks via the
relative pressure jump $\displaystyle\frac{\Delta p}{p_0}$,
where $\Delta p=p_1-p_0$, and $p_0$, $p_1$ are the values of
pressure before and after the shocks. Using formulae [\ref{Samarskii}]:

$$
\left(\frac{\Delta p}{p_0}\right)_1=
\frac{2\gamma}{\gamma+1}\left(\frac{(D_1-V)^2}{c_s^2}-1\right)\,,
$$

$$
\left(\frac{\Delta p}{p_0}\right)_2=
\frac{2\gamma}{\gamma+1}\left(\frac{(D_2-V_1)^2}{c_s^2}-1\right)\,,
$$
and substituting $\gamma=\slantfrac53$, $c_s=5$~km/s,
$V=25$~km/s, $V_1=40$~km/s, $D_1=37$~km/s,
$D_2=28$~km/s we obtain $\displaystyle\left(\frac{\Delta
p}{p_0}\right)_1= \displaystyle\left(\frac{\Delta
p}{p_0}\right)_2=6$ (the equal strength of two shocks is obvious
due to the symmetry of the problem in an appropriate coordinate
system).

\begin{figure}
\centerline{\hbox{\psfig{figure=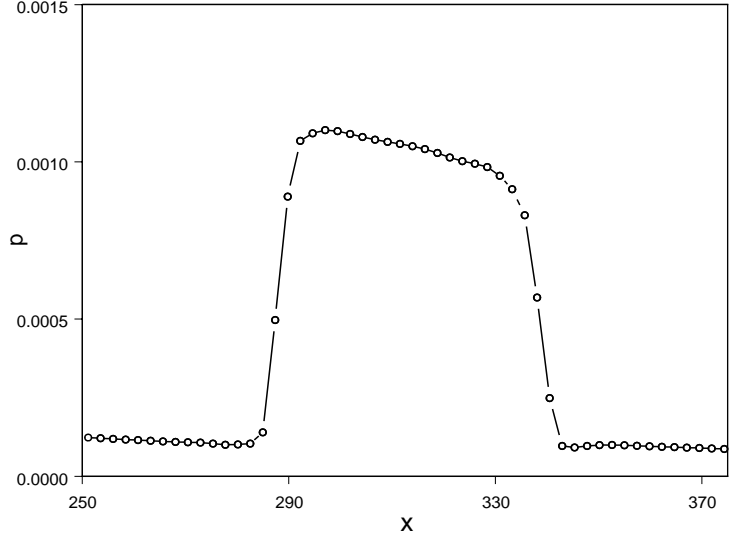,width=10cm}}}
\caption{\small The pressure profile along the $x$ axis in the area near the double
shock wave for the moment $0.073P_{orb}$ after the wind velocity change
(shown in bottom panel of Fig.~\ref{double_shock}). Axes in the figure: $x$-axis --
distance -- in solar radii,
$y$-axis -- pressure -- in relative
units.}
\label{Pressure}
\end{figure}

Figure~\ref{Pressure} illustrates the pressure change in the area near the bow
shocks.  Here we can see that in the area between the two shocks the
jump in pressure is in accordance with the above estimates.

When the first of shocks reaches the bow shock located in front
of the accretor (after $0.092P_{orb}$ in the case of $V_1=40$~km/s)
the process of the flow rearrangement near the accretor starts. The
coming shocks begin to interact with the matter of the disk and
crush it forcing the most part of the matter composing the disk to
fall on the accretor's surface.

\begin{figure}
\vspace*{3mm}
\centerline{\hbox{\psfig{figure=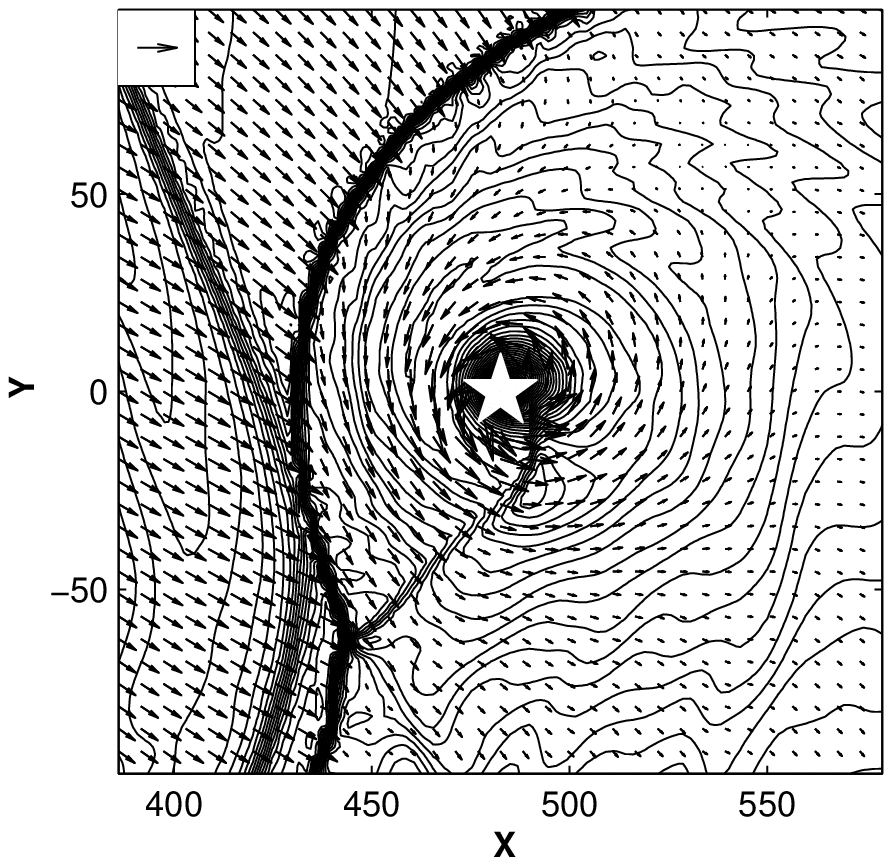,width=7cm}}\hbox{\psfig{figure=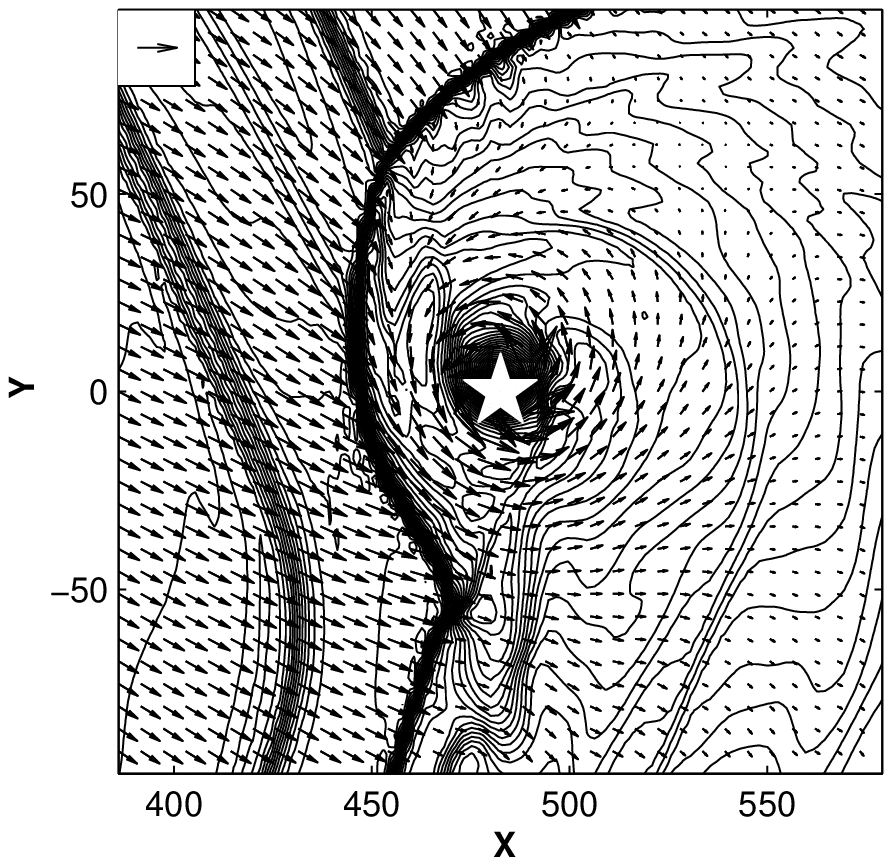,width=7cm}}}
\vspace*{3mm}
\centerline{\hbox{\psfig{figure=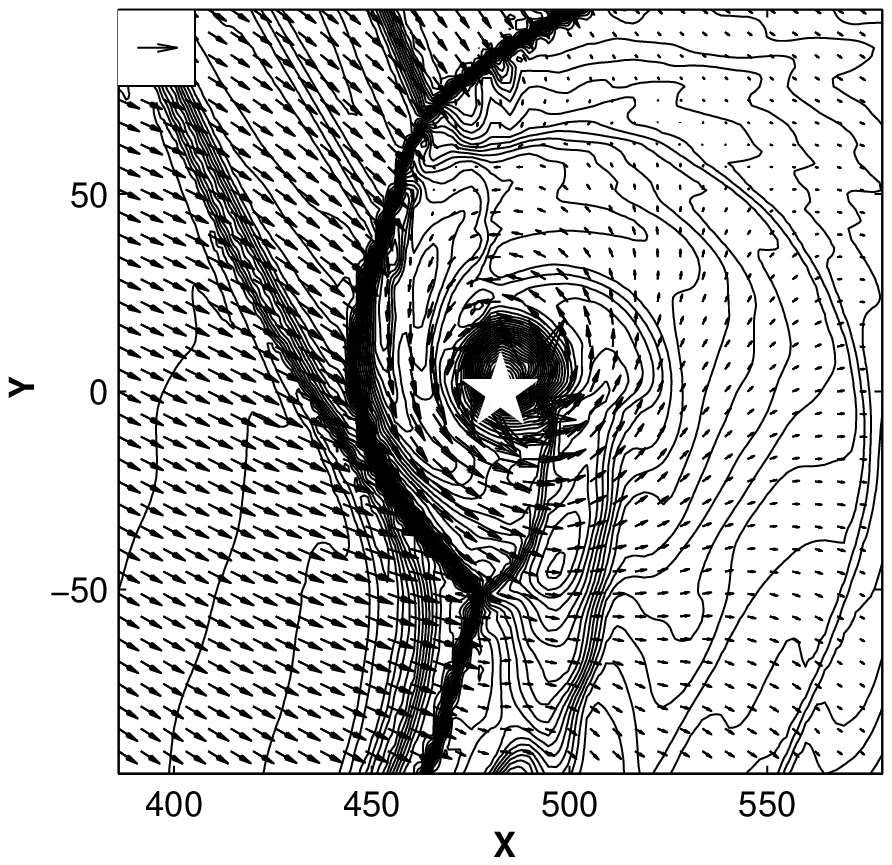,width=7cm}}\hbox{\psfig{figure=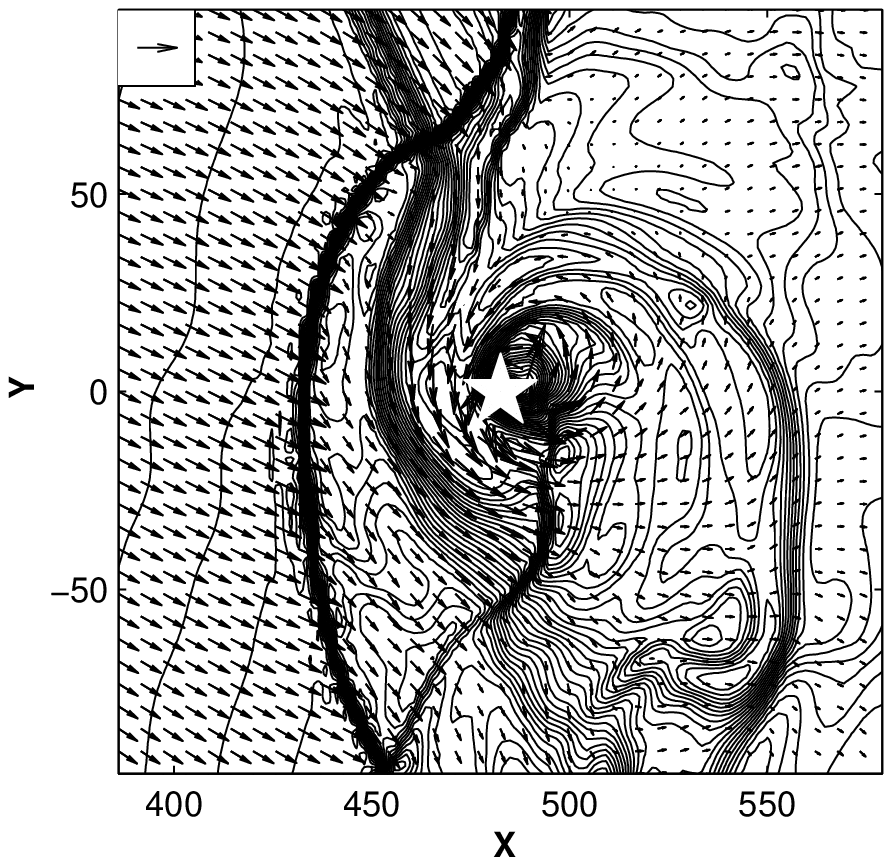,width=7cm}}}
\vspace*{5mm}
\caption{\small Density field and velocity vectors in the area near
the accretor for solution with wind velocity change from $V=25$~km/s to
$V_1=40$~km/s. The arrow in the left upper corner corresponds to the
velocity of 100~km/s. All other designations are the same as in
Fig.~\ref{regimes}. Four moments are presented: left top panel -- $0.092P_{orb}$;
right top -- $0.110P_{orb}$; left bottom -- $0.115P_{orb}$;
right bottom -- $0.137P_{orb}$ after the wind velocity change.}
\label{interact}
\end{figure}

The set of panels in Figure~\ref{interact} presents the morphology of the gas
flow in the vicinity of the accretor for different moments at early
stages of the flow rearrangement. Here the area
$[0.8A\ldots1.2A]\times[-0.2A\ldots 0.2A]$ (or $80\times80$ grid
cells) is presented. After the time of about $0.09P_{orb}$ ($\sim$70
days) after the wind velocity change the first of the two shocks
formed reaches the vicinity of the accretor, namely the bow shock in
front of it (Fig.~\ref{interact} -- left top panel). The first wave makes the bow
shock to come closer to the accretor (see Fig.~\ref{interact} -- right top panel)
but the disk still exists. Then ( $\sim14 $ days later) the second wave
comes (see Fig.~\ref{interact} -- left bottom panel) and destroys the disk
(see Fig.~\ref{interact} -- right bottom panel).

\begin{figure}
\centerline{\hbox{\psfig{figure=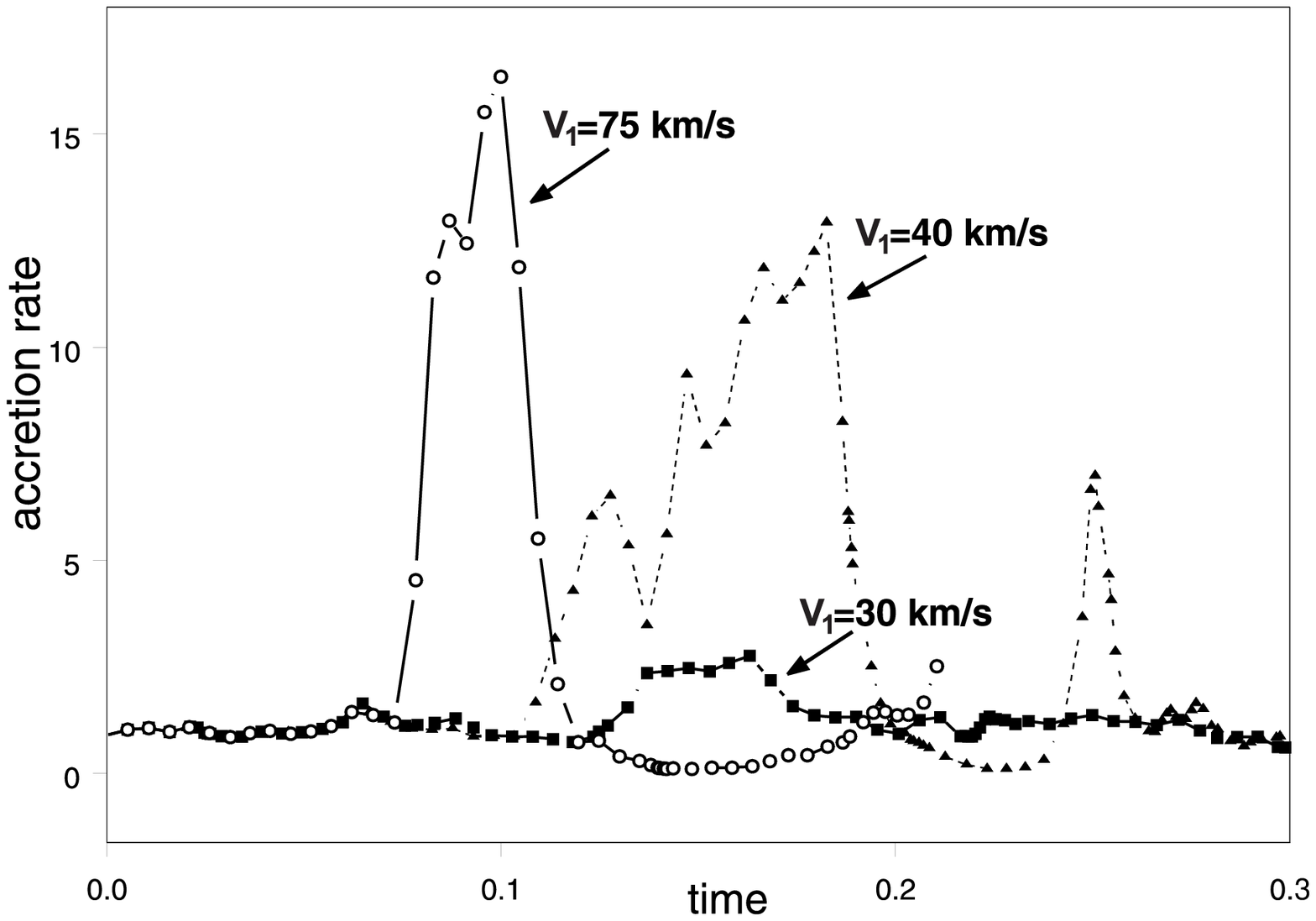,width=10cm}}}
\caption{\small The accretion rate change with time for the cases of wind velocity change
from $V=25$ to $V_1=$30, 40 and 75 km/s. Time is given in units of the orbital period
($P_{orb}$=758 days), the accretion rate is in percents of the matter that leaves
donor's surface.}
\label{Accr_rate}
\end{figure}

The accretion rate during the flow rearrangement undergoes
significant changes. The accretion rate behaviour during the process
of the flow rearrangement is presented in Figure~\ref{Accr_rate} for different
values of the final wind velocity.

We can see that when the final wind velocity $V_1$ equals 30 km/s there is
no significant increase of the accretion rate while some small
variations exist. These variations are probably due to disturbances
of the accretion disk by the shock waves formed after the velocity increasing
that are not strong enough to crush it. The situation is principally
different for other cases -- here the jump in the accretion rate
takes place. For example, when the final wind velocity is equal to
40 km/s the accretion rate jumps in more than 14 times and in case
of 75~km/s -- in more than 16 times. This abrupt increase of the
accretion rate is caused by the destruction of the accretion disk.

It should be mentioned that the study of the transition period is
limited in the framework of this model. After the accretion rate
jump the accepted boundary conditions on the accretor change and the
model used doesn't describe the real physical situation any more.
Correspondingly, the presented results of calculations of the flow
rearrangement period are correct only at first stages of the process.

According to the estimates based on observations [\ref{FC95}] in the
case of Z And the donor's wind velocity is $25-40$ km/s when the
system is at quiescence. These values of the wind velocity are close
to the critical value 35 km/s dividing different accretion regimes.
If we suppose the presence of a minor variations of the wind
velocity caused, e.g., by the typical giant's activity [\ref{MK92}]
the wind velocity can, at some moment, become greater than the
critical value. As the consequence the change of the flow structure
and of the accretion regime will occur followed by the abrupt jump of the
accretion rate in the transition period.

\bigskip

\section{Conclusions}

\bigskip

The 2D numerical investigation of the gas flow structure in the
interacting binary systems with the mass exchange via stellar wind
(and namely for their subclass -- symbiotic systems) has been
carried out using the Roe-Oscher scheme. The modelling has been
carried out for different values of the wind velocity. The effect of
the velocity change on the system's behavior was of special
interest.

It is shown that after the wind velocity increase the collision of
two codirectional flows results in the formation of two shock waves
that begin to propagate from the donor's surface. These shocks can
lead to the significant changes in the flow structure near the
accreting component and namely to the accretion regime change.

These results were used as the ground for the new mechanism
providing the change of the accretion rate and explaining the nature
of outbursts in symbiotic stars. The point of this mechanism is in
the abrupt change of the flow structure near the accreting component
as a result of minor fluctuations of the wind from the
mass-losing component. These changes come out in the accretion
regime change, namely, in transition from the disk accretion to the
accretion from the flow when the wind velocity increases. The
process of the flow rearrangement in such cases is followed by the
jump in the accretion rate.

Results of modelling of the flow structure in the classical
representative of the studied type of binary systems Z~And show that
in the quiescent state characterized by the wind velocity from 25 to
40~km/s the steady accretion disk forms in the system. In the given
regime the accretion efficiency is $\sim $1\% of the donor's mass
loss. If the wind velocity increases and becomes greater than
35~km/s the disk disappears and the cone shock wave forms. In
stationary solutions for high-velocity regimes ($35-75$ km/s) the
accretion rate is slightly higher and equals $\sim $2\% of the
donor's mass loss. But during the flow rearrangement when the wind
with increased velocity reaches disk and crushes it the accretion
rate increases in dozens times!

The analysis of the observations of symbiotic stars during the
active stages shows that the most part of the registered
manifestations can be explained in terms of the drop-off of the
optically thick shell by the accretor [\ref{FC95}]. The suggested
mechanism provides the accretion rate jump and corresponding
increase of the energy release rate large enough to provide the
drop-off of the part of the gas from the accretor. Therefore it leads
to the appearance of the flow features that can explain the
observations.

\section{Acknowledgements}
\bigskip

This work was partly supported by the Russian Foundation for Basic
Research (projects \N\N~02-02-16088, 02-02-17642, 00-01-00392), RF
President grant 00-15-967221, Federal Programs "Mathematical Modelling"
and "Astronomy" as well as by INTAS (grant \N~00-491).

\section*{References}

\begin{enumerate}

\item
\label{Kniga} A. A. Boyarchuk, D. V. Bisikalo,
O. A. Kuznetsov and V. M. Chechetkin, Mass Transfer
in Close Binary Stars, Taylor\&Francis (2002).

\item
\label{Boyar92} A. A. Boyarchuk, in {\it Variable Stars and
Galaxies}, Ed. by B. Warner (ASP Conference Series, 1992),
{\bf 30}, pp. 325--338.

\item
\label{MK92} J. Miko\l ajewska and S. J. Kenyon, Monthly
Notices Roy. Astron. Soc. {\bf 256}, 177 (1992).

\item
\label{FC95} T. Fernandez-Castro, R. Gonzales-Riestra, A.
Cassatella, A. R. Taylor and E. R. Seaquist, Astrophys. J.  {\bf 442}, 366 (1995).

\item
\label{TU} A. V. Tutukov and L. R. Yungelson, Astrofiz. {\bf
12}, 521 (1976).

\item
\label{Zytkow}
B. Paczy\'nski and A. \.Zytkow, Astrophys. J. {\bf 222},
604 (1978).

\item
\label{Rudak}
B. Paczy\'nski and B. Rudak, Astron. \& Astrophys. {\bf 82},
349 (1980).

\item
\label{Sion}
E. M. Sion, M. J. Acierno, and S. Tomczyk, Astrophys. J. {\bf
230}, 832 (1979).

\item
\label{Iben}
I. Iben, Jr., Astrophys. J., {\bf 259}, 244 (1982).

\item
\label{Fuji}
M. Y. Fujimoto, Astrophys. J. {\bf 257}, 767 (1982).

\item
\label{IbenTu}
I. Iben, Jr. and A. V. Tutukov, Astrophys. J. {\bf 342},
430 (1989).

\item
\label{Friedjung93} M. Friedjung, in {\it Cataclysmic Variables
and Related Objects}, Ed. by M. Hack and C. La Douse (US Gov.
Printing Office, Washington, 1993), pp. 647--655.

\item
\label{Kenyon83} S. J. Kenyon and J. S. Gallagher,
Astron. J. {\bf 88}, 666 (1983).

\item
\label{GS}
G. S. Bisnovatyi-Kogan, Ya. M. Kazhdan, A. A. Klypin, A. E. Lutskii
and N. I. Shakura, Astron.  Zh. {\bf 56}, 359 (1979)
 [Sov.  Astron. {\bf 23}, 201 (1979)].

\item
\label{Dima94}
D. V. Bisikalo, A. A. Boyarchuk, O. A. Kuznetsov and V. M. Chechetkin,
Astron.  Zh. {\bf 71}, 560 (1994) [Astron. Reports {\bf 38}, 494
(1994)].

\item
\label{Dima96}
D. V. Bisikalo, A. A. Boyarchuk, O. A. Kuznetsov and V. M. Chechetkin,
Astron.  Zh. {\bf 73}, 727 (1996) [Astron. Reports {\bf 40}, 662
(1996)].

\item
\label{Lena} E. Kilpio, Odessa Astron. Publ. {\bf 14}, 41
(2001).

\item
\label{Theuns1}
T. Theuns and A. Jorissen, Monthly Not. Royal Astron. Soc. {\bf
265}, 946 (1993).

\item
\label{Theuns2}
T. Theuns, H. M. J. Boffin, and A. Jorissen, Monthly Not. Royal
Astron. Soc. {\bf 280}, 1264 (1996).

\item
\label{BellLin}
K. R. Bell and D. N. Lin, Astrophys. J. {\bf 427}, 987 (1994).

\item
\label{FC88} T. Fernandez-Castro, A. Cassatella, A. Gimenez
and R. Viotti, Astrophys. J. {\bf 324}, 1016 (1988).

\item
\label{Roe} P. L. Roe, Ann. Rev. Fluid Mech. {\bf 18}, 37 (1986).

\item
\label{Osher} S. Chakravarthy and S. Osher, AIAA Pap N 85--0363
(1985).

\item
\label{AZh2002} D. V. Bisikalo,  A. A. Boyarchuk, E. Yu. Kilpio and O. A. Kuznetsov
Astron.  Zh. {\bf 79},(2002) (in press)

\item
\label{Samarskii} A. A. Samarskii and Yu. P. Popov, Finite-Difference Methods of
Solution of Gas Dynamics, Moscow, Nauka Academic Press (in Russian) (1980).

\end{enumerate}

\clearpage

\end{document}